\input harvmac

\def\xxx#1 {{hep-th/#1}}
\def\lr { \lref}
\def\npb#1(#2)#3 { Nucl. Phys. {\bf B#1} (#2) #3 }
\def\rep#1(#2)#3 { Phys. Rept.{\bf #1} (#2) #3 }
\def\plb#1(#2)#3{Phys. Lett. {\bf #1B} (#2) #3}
\def\prl#1(#2)#3{Phys. Rev. Lett.{\bf #1} (#2) #3}
\def\physrev#1(#2)#3{Phys. Rev. {\bf D#1} (#2) #3}
\def\ap#1(#2)#3{Ann. Phys. {\bf #1} (#2) #3}
\def\rmp#1(#2)#3{Rev. Mod. Phys. {\bf #1} (#2) #3}
\def\cmp#1(#2)#3{Comm. Math. Phys. {\bf #1} (#2) #3}
\def\mpl#1(#2)#3{Mod. Phys. Lett. {\bf #1} (#2) #3}
\def\ijmp#1(#2)#3{Int. J. Mod. Phys. {\bf A#1} (#2) #3}
\def\mpla#1(#2)#3{Mod. Phys. Lett. {\bf A#1} (#2) #3}
\def\jhep#1(#2)#3{JHEP {\bf  #1} (#2) #3}

\parindent 25pt
\overfullrule=0pt
\tolerance=10000

\def\half{{\textstyle {1 \over 2}}}



\lr\costas{C. Bachas, C. Fabre, E. Kiritsis, N.A. Obers and P.
Vanhove, {\it Heterotic / type I duality and D-brane instantons},
\npb509(1998)33, hep-th/9707126.}
 \lr\kiritsis{E. Kiritsis and
N.A. Obers, {\it  Heterotic type I duality in D $<$ 10-dimensions,
threshold corrections and D instantons}, \jhep10(1997)004,
hep-th/9709058.} 
\lr\polchinskia{J. Polchinski and E. Witten, {\it
Evidence for heterotic - type I string duality}, \npb460(1996)525,
\xxx9510169.}
 \lr\greenguta{M.B. Green and M. Gutperle, {D-particle bound states and the
D-instanton measure}, \jhep01(1998)005, \xxx9711107;  {\it D instanton
partition functions}, 
\physrev58(1998)046007,
\xxx9804123.}
 \lr\bgs{C. Bachas, M.B.
Green and A. Schwimmer, {\it (8,0) quantum mechanics and symmetry
enhancement in type I' superstrings}, \jhep01(1998)006,
hep-th/9712086.} 
\lr\kachru{Kachru and Silverstein, {\it On gauge
bosons in the matrix model approach to M theory}, \plb396(1997)70,
\xxx9612162.}
\lr\dixon{L. Dixon,
V. Kaplunovski and J. Louis, {\it Moduli dependence of string loop
corrections to gauge coupling constants}, \npb355(1991)649.}
\lr\yi{P.Yi, {\it Witten index and threshold bound states of
    D-branes}, \npb505(1997)307,  \xxx9704098.} 
\lr\sethi{S. Sethi and M. Stern, {\it D-brane bound states redux},
  \cmp194(1997)675, \xxx9705046.}
\lr\nikita{G. Moore, N. Nekrasov and S. Shatashvili, {\it D particle
    bound states and generalized instantons}, \xxx9803265.}
\lr\lerche{W. Lerche and S. Stieberger,
{\it Prepotential, mirror map and F theory on K3}, \xxx9804176; W.
Lerche, S. Stieberger
  and N. Warner, {\it Quartic gauge couplings from K3 geometry},  \xxx9811228.}
\lr\stephana{K. Foerger and S. Stieberger, {\it Higher derivative
couplings and heterotic type I duality in eight-dimensions},
\xxx9901020.}
\lr\stephanb{P. Mayr and S. Stieberger, {\it Threshold corrections to
gauge couplings in orbifold compactifications}, \npb407(1993)725,
\xxx9303017.}

 \lr\ellis{J. Ellis, P. Jetzer, L. Mizrachi, {\it One loop string corrections
 to the effective field theory}, \npb303(1988)1.}
\lr\lowe{D. Lowe, {\it Bound states of type I-prime D particles
and enhanced gauge symmetry}, \npb501(1997)134, \xxx9702006.}
 \lr\ferr{U.H. Danielsson and G.~Ferretti,
{ \it The Heterotic life of the D particle}, \ijmp12(1997)4581,
\xxx9610082.}
 \lr\ellgenus{W. Lerche, B. Nilsson and A. Schellekens, {\it Heterotic String
  Loop Calculation Of The Anomaly Cancelling Term}, \npb289(1987)609;
  W. Lerche, B. Nilsson, A. Schellekens and N. Warner, {\it Anomaly Cancelling
  Terms From The Elliptic Genus}, 
  \npb299(1988)91.}
  \lr\lercheb{W. Lerche, {\it Elliptic Index And Superstring Effective
      Actions}, \npb308(1988)102.} 
\lr\horava{P. Horava and E. Witten, {\it Heterotic and type I
string dynamics from eleven-dimensions}, \npb460(1996)506,
\xxx9510209.}
\lr\wittena{E. Witten, {\it 
String theory dynamics in various dimensions}, 
\npb443(1995)85,
hep-th/9503124.}
\lr\greengutb{M.B. Green and M. Gutperle,
{\it Effects of D instantons},
\npb498(1997)195, hep-th/9701093.}
\lr\banksa{T. Banks, N. Seiberg and E.~Silverstein, {\it 
Zero and one-dimensional probes with N=8 supersymmetry}, 
\plb401(1997)30, \xxx9703052.}

\lr\banksmotl{T. Banks and L. Motl, {\it Heterotic Strings from
Matrices}, \jhep12(1887)004, \xxx9703218.}

\lr\rey{Soo-Jong Rey, {\it Heterotic M(atrix) Strings and
Their Interactions}, \npb502(1997)170,  \xxx9704158.}

\lr\hor{P. Horava, {\it Matrix Theory and Heterotic Strings
on Tori}, Nucl. Phys. {\bf B505} (1997) 84, \xxx9705055.}

\lr\kabat{ D. Kabat and S.-J. Rey, {\it Wilson Lines and T-Duality
in Heterotic Matrix Theory}, \npb508(1997)535, \xxx9707099.}

\lr\nilles{D. Matalliotakis, H-P. Nilles and S. Theisen,
{\it Matching the BPS Spectra of Heterotic-Type I-Type I' Strings},
\plb421(1998)169, 
\xxx9710247.}

\lr\berg{O. Bergman, M. R. Gaberdiel and  G. Lifschytz,
{\it  String Creation and Heterotic-Type I' Duality},
\npb524(1998)524, \xxx9711098.}

\lr\gava{M. Bianchi, E. Gava, F. Morales and K.S. Narain, {\it D
strings in unconventional type I vacuum configurations}, \xxx9811013.}

\lr\dhar{J.R. David, A. Dhar and  G. Mandal, {\it Probing Type I' String Theory Using D0 and D4-Branes}, \plb415(1997)135, \xxx9707132.} 

 \noblackbox 
\baselineskip 18pt plus 2pt minus 2pt
\Title{\vbox{\baselineskip12pt
\hbox{hep-th/9903010}
\hbox{PUPT-1841}
}}
{\vbox{
\centerline{A note on heterotic/type I$^\prime$ duality}
 \centerline{and D0 brane quantum mechanics} }}
\centerline{Michael Gutperle\foot{email:
gutperle@feynman.princeton.edu}} 
\medskip
\centerline{Department of Physics, Princeton University, Princeton NJ
08554, USA} \bigskip

\medskip
\centerline{{\bf Abstract }}
In this note a simple  calculation of one loop threshold corrections for the
$SO(32)$ heterotic string is performed. In particular the
compactification on $T^2$ with a  Wilson line breaking the gauge group
to $SO(16)\times SO(16)$ is considered. Using heterotic type I
duality, these corrections can be related to quantities appearing in the
quantum mechanics of type $I^\prime$ D0 particles. 
\noblackbox
\baselineskip 16pt plus 2pt minus 2pt

\Date{February 1999}
\vfill\eject

\newsec{Introduction}
The quantum mechanics of D0 branes and coming with it 
 the question of existence of 
bound states of D0 branes is important for various string 
dualities. M-theory/type IIA duality implies that there is a single  bound
state of $N$ D0 branes for any $N$  corresponding to  Kaluza-Klein modes on 
the M-theory circle \wittena. The relevant index for the case of two D0
branes was computed in \sethi,\yi. In this calculation the index is
split into a 
bulk and boundary term which in turn is expressed as a zero
dimensional matrix integral, whose value for arbitrary $N$ was
conjectured in \greenguta\ using results from \greengutb. The bulk term
of the index was    calculated directly  in \nikita\  using methods of
topological field theory.  

Another interesting example  of quantum mechanics of D0 particles appears
in   the so called type $I^\prime$ theory, which describes D0 
particles in the presence of D8 branes and orientifold O8 planes
\ferr\banksa\bgs\berg. This theory is important for the matrix theory
formulation  of the heterotic string \banksmotl\rey\hor\kabat\dhar.
Type I$^\prime$ string theory is defined as the IIA  orientifold 
on $S^1/Z_2$ which is T-dual to type I theory.   There are two
orientifold  eight planes at the two ends of the interval  and
sixteen D8 branes in between. If eight D8 branes are on top of
each orientifold plane, the gauge symmetry is given by
$SO(16)\times SO(16)$. The strong coupling limit of this system is
given by M-theory on $S^1/Z_2$ \horava. The states which
are not present in perturbative type I$^\prime$ spectrum, but
which are needed to fill out the $E_8\times  E_8$ multiplets, are
given by bound states of D0-particles \kachru\lowe. Hence
we have to look for bound states of D0-particles transforming in
the ${\bf 128}$ and ${\bf 120}$ of $SO(16)$.

It is interesting to find the description of these states in the
heterotic $SO(32)$ theory where they are perturbatively realized.
These states are BPS states with  $N_R=0(1/2)$ for the R(NS) sector.  
The mass and level matching \berg\kachru\nilles\ conditions become
\eqn\levelm{{p_R^2\over 2}= {p_L^2\over 2}+N_L-1,\quad m^2= p_R^2,}
where the right moving momenta in $\Gamma_{17,1}$ are given by
\eqn\momnhet{\eqalign{p_L&=\big(P+Yn,{m-1/2 Y^2n- YP\over 2 R}-nR\big)\cr
p_R&={m-1/2 Y^2n- YP\over 2 R}+nR
,}}
Here  $Y$ is a Wilson line along $S^1$ and $P$ are momenta in $SO(32)$
lattice and $m$ and $N$ are the momentum and the winding along $S^1$
respectively.
With a Wilson line  given by $Y=(0^8,(1/2)^8)$, $SO(32)$ is
broken to $SO(16)\times SO(16)$ and the analysis in \kachru\ shows that
the states with $N_L=0$ and even $n$ lie in the $({\bf 120,1})+({\bf
  1,120})$ of 
$SO(16)\times SO(16)$ whereas the states with odd $n$ lie in $({\bf
128,1})+(\bf {1,128})$
\newsec{Heterotic one loop thresholds}
The duality of the $SO(32)$ heterotic and type I  strings in ten
dimensions makes it possible to calculate  some nonperturbative
effects on the type I side  due to Euclidean D-branes exactly by a
one loop calculation on the heterotic side. The simplest case in
which such a calculation is possible arises for  the heterotic
string compactified on a two torus $T^2$
\costas\kiritsis\lerche\stephana\gava. Worldsheet instantons on the
heterotic side  get 
mapped to wrapped Euclidean D-branes on the torus, which provide
D-instanton effects in eight dimensions.

There are one loop heterotic thresholds \ellis\lercheb\ which are
BPS-saturated and related by supersymmetry to anomaly canceling
terms \ellgenus\ and therefore presumably exact at one loop. The one loop
integrals involved are almost holomorphic since only BPS-states
run in the loop and hence the loop integrals  can be calculated exactly.  
For the $SO(32)$ heterotic string the relevant  
loop amplitude  for gravitational thresholds with Wilson lines 
is given by 
\eqn\loopone{I_d= -{\cal N} (2\pi)^d  \int_F{d^2\tau\over
      \tau_2^{2-d/2}} \Gamma_{d,d+16}{ A}(R,\tau),}
with ${\cal N}= V^{(10-d)}/  (2^{10}\pi^6)$.  The lattice function
$\Gamma_{d,d+16}(G,B,Y)$ 
is given by
\eqn\looptwo{\eqalign{\Gamma_{d,d+16}&={\sqrt{det(G)}\over
    \tau_2^{d/2}} \sum_{m^i,n^i} e^{-{\pi\over
      \tau_2}(G+B)_{ij}(n+m\tau)^i(n+m\bar{\tau})^j}\cr
&\times \sum_{a,b=0,1}\prod_{k=1}^{16} e^{-i\pi(m^in^jY_i^k Y_j^k+b m^iY_i^k)}
\theta\left[\matrix{a+2m^l Y_l^k\cr b+2n^l Y_l^k}\right](0,\tau).}}
Here $Y_i^k$ $i=1,\cdots,16$, $k=1,\cdots,d$ parameterize the Wilson
lines around the cycles  of $T^d$. The almost holomorphic $A$ is given
by
\eqn\loopthree{A(R,\tau)= {1\over 2^7 3^2 5}{E_4(\tau)\over
      \eta^{24}(\tau)} t_8 \tr R^4+ {1\over 2^9 3^2
      }{\hat{E}_2^2(\tau)
\over
      \eta^{24}(\tau)} t_8 (\tr R^2)^2,}
where $E_{2n}(\tau)$ are the Eisenstein modular forms of weight
$2n$.
\newsec{Two torus compactification with Wilson lines}

We are interested in the $T^2$ compactification with the K\"ahler
and complex structure modulus $T,U$. The $SO(32)$
gauge symmetry will be broken  to $SO(16)\times SO(16)$ by
introducing Wilson lines on the two torus of the following form
\eqn\wilsonl{Y_i^1=(0^8,\half^8), Y_i^2=(0^8,0^8). } This choice of
Wilson lines corresponds in the type I$^\prime$ picture to eight
D8 branes sitting on top of each of the two  O8 planes, canceling
dilaton and Ramond-Ramond sources locally.

 We want to calculate  the one
loop thresholds of the form $t_8 \tr(R^4)$, $t_8\tr(F^4)_{1}$ and
$t_8(\tr(F^2)_{1})^2$ in the presence of this  Wilson line. The
subscript on the field  strength  in the second and third 
term indicates that the trace is taken over the first $SO(16)$ factor.
  The integrals that will appear in this
calculations are of the following form
\eqn\oneloopfour{I_{Q}=\int_F{d^2\tau\over
      \tau_2} \sum_{A} {T_2\over \tau_2} \exp\Big\{ 2\pi i T
      \det{A}-{\pi T_2\over \tau_2 U_2}\big| (1 U) A
          \Big(\matrix{\tau \cr 1}\Big)\big|^2\Big\}  Q C(Y,A).}
Here the matrix A is given by  2$\times$2 matrices with integer
entries
\eqn\amat{A= \pmatrix{m_1&  n_1\cr m_2 &n_2},\quad m_1,m_2,n_1,n_2\in Z,}
 and $C(Y,A)$ is the partition function of the $SO(32)$ lattice which 
in general depends on the Wilson line $Y$ and the matrix
 $A$ and is given by
\eqn\intlat{\eqalign{C(Y,A)&=  \sum_{a,b=0,1}\prod_{k=1}^{16}
e^{-i\pi(m^in^jY_i^k Y_j^k+b n^iY_i^k)} \theta\left[\matrix{a+2m^l
Y_l^k\cr b+2n^l Y_l^k}\right](0,\tau) \cr &=
\sum_{a,b}\theta^8\left[\matrix{a\cr
b}\right](0,\tau)\theta^8\left[\matrix{a+m_1\cr
b+n_1}\right](0,\tau).}}  We
introduced the standard notation for the theta functions
\eqn\thetnot{\theta\left[\matrix{1\cr
      1}\right]=\theta_1,\quad\theta\left[\matrix{1\cr
      0}\right]=\theta_2,\quad \theta\left[\matrix{0\cr
      0}\right]=\theta_3,\quad\theta\left[\matrix{0\cr
      1}\right]=\theta_4,\quad.}
The form of the operator $Q$ in
\oneloopfour\ depends on the threshold in question. For gravitational
thresholds $t_8\tr(R^4)$ and $t_8 (\tr(R^2))^2$ $Q$ is independent of
the spin structures and $A$ in \intlat\ and is given by \loopthree.

The operator $Q$ for $\tr(F^4)$ and $(\tr(F^2))^2$ can be found by
'gauging' \intlat \ellis. The Wilson line \wilsonl\ breaks the gauge
group to $SO(16)\times SO(16)$ and the thirty two free fermions of the
$SO(32)$ lattice are split into two sets of sixteen in \intlat.
The result depends on the spin structures $[a,b]$ for 
the sixteen fermions which are associated with the first $SO(16)$ in
\intlat. For the $\tr(F^4)$ threshold the operators are given by 
\eqn\qresult{\eqalign{Q_{\tr(F^4)}\left[\matrix{1\cr
        0}\right](\tau)&=-{1\over 2^8 3} 
  \theta_3^4\theta_4^4(\tau),\cr
 Q_{\tr(F^4)}\left[\matrix{0\cr 0}\right](\tau)&={1\over
     2^8 3} \theta_2^4\theta_4^4(\tau),\cr 
Q_{\tr(F^4)}\left[\matrix{0\cr   1}\right](\tau)&=-{1\over   2^8 3} 
\theta_2^4\theta_3^4(\tau), }} 
whereas for the $(\tr(F^2))^2$ threshold the operator is given by 
\eqn\trftwoq{\eqalign{Q_{(\tr(F^2))^2}\left[\matrix{1\cr
0}\right](\tau)&={1\over 2^{10}3^2}\big(e_2(\tau)+\hat{E}_2(\tau)\big)^2,\cr
 Q_{(\tr(F^2))^2}\left[\matrix{0\cr
0}\right](\tau)&={1\over 2^{10}3^2}\big(e_3(\tau)+\hat{E}_2(\tau)\big)^2,\cr
Q_{(\tr(F^2))^2}\left[\matrix{0\cr
1}\right](\tau)&={1\over 2^{10}3^2}\big(e_4(\tau)+\hat{E}_2(\tau)\big)^2.}}
 Where the following
notation has been introduced \eqn\edefin{e_2=
\theta_3^4+\theta_4^4,\quad e_3= \theta_2^4-\theta_4^4,\quad e_4=
-\theta_2^4-\theta_3^4,}
and $\hat{E}_2$ is the nohomolomorphic (but modular) Eisenstein
function of weight 2.

\newsec{Evaluation of integral}
The integral \oneloopfour\ can be evaluated  using the method
of orbits \dixon. In the present context this technique was
discussed in \costas\kiritsis\ and in \lerche, where type I
thresholds  with certain Wilson lines present were evaluated using
results from \stephanb.  Without 
Wilson lines it is straightforward to show that 
 under the modular $SL(2,Z)$ transformations $\tilde{\tau}=
 (a\tau+b)/(c\tau+d)$ with $a,b,c,d\in Z$,$ad-bc=1$ 
   \eqn\stmat{{1\over \tilde{\tau}_2 }\big| (1 U) { A}
          \pmatrix{\tilde{\tau} \cr 1}\big|^2= {1\over {\tau}_2 }\big|
          (1 U)  {A} \pmatrix{a&b\cr c&d} 
          \pmatrix{{\tau} \cr 1}\big|^2.}
 The summation over all integer matrices  matrices ${ A}$ can then 
 replaced  by the summation over all equivalence classes of
 $SL(2,Z)$ orbits.  There are 
three different cases, the trivial orbit ${A}=0$, the degenerate
orbit $\det({A})=0$ and the non degenerate orbit $\det({A})\neq 0$

In the following we will consider only  the non degenerate orbit, where
the  fundamental ${\cal F}$ is unfolded into the double cover of the
upper half plane ${\cal H}$. The non degenerate $SL(2,Z)$ orbits fall
into the following equivalence classes  
\eqn\orbitrf{{ A}=\pm \pmatrix{k&j\cr 0& p},\quad k>0, 0\leq j<k, p\in Z.}
When Wilson lines are present, matters are more complicated but using
the well known transformation properties of the theta functions under
$\tau\to \tau+1,\tau\to -1/\tau$ is is easy to see that for both
$Q_{\tr(F)^4}$ \qresult\ and $Q_{(\tr(F)^2)^2}$ \trftwoq, $QC(Y,A)$
defined in \intlat\ behaves in the following way
\eqn\trafq{QC(Y,{A})({a\tau+b\over c\tau+d})=QC(Y,{ A}
  \pmatrix{a&b\cr c&d} )(\tau).} 
Hence the method of orbits can be used to unfold the integral. For
the non degenerate orbit we get
\eqn\unfoldedint{I_{nd}=\int_{H}{d^2\tau\over
      \tau_2} \sum_{\quad k>0, 0\leq j<k, p\in Z}{T_2\over \tau_2} 
       \exp\Big\{ 2\pi i kpT
      -{\pi T_2\over \tau_2 U_2}\big| k\tau +j +pU|^2\Big\} Q
      C(Y,\pmatrix{k&j\cr0&p})(\tau).} 
In order to evaluate \unfoldedint\ it is convenient  to  split the summation over
equivalence classes ${ A}$ in \orbitrf\ into four seperate sectors  ${
  A}^{(i)},i=1,\cdots,4$.
\eqn\amas{\eqalign{A^{(1)}&= \pmatrix{2\tilde{k} & 2\tilde{j} \cr 0
      &p},\quad\quad 0\leq 2\tilde{j}<2\tilde{k},\cr
A^{(2)}&=\pmatrix{2\tilde{k}+1&  2\tilde{j}\cr 0 &p},\quad 0\leq
2\tilde{j}<2\tilde{k}+1,\cr 
 A^{(3)}&=\pmatrix{2\tilde{k}&
    2\tilde{j}+1\cr 0 & p },\quad 0\leq 2\tilde{j}+1<2\tilde{k},\cr
\quad A^{(4)}&= \pmatrix{2\tilde{k}+1&  2\tilde{j}+1\cr 0
    &p}\quad 0\leq 2\tilde{j}+1<2\tilde{k}+1.}}
The expansion of $Q C(Y,A^{(i)})$ appearing in \unfoldedint\ in powers
of $q=\exp(2\pi i \tau)$ and powers of $1/\tau_2$ is given by 
\eqn\expanq{Q C(Y,A^{(i)})(\tau)= \sum_{n\geq -1, r\geq 0} 
    c^{(i)}_{n,r} \;{1\over \tau_2^r}q^n.}
The integral \unfoldedint\ is then of the form $I_{n,r}$ defined in appendix.
Such integrals were evaluated in \costas\kiritsis\ and the
main results are reviewed in the appendix for completeness. 

The terms of order $1/q$ in \expanq\ are problematic for the type
I$^\prime$ interpretation as discussed in section 6.  
For all $QC(A^{(i)})$  which will be considered later it turns out that only
  the $A^{(1)}$
and $A^{(3)}$ sector contribute  terms of order $1/q$ in the
integral. In addition we shall find that 
$c^{(1)}_{-1,r}=c^{(3)}_{-1,r}$. In this case the summation over $\tilde{j}$
of the two terms can be combined giving $\sum_{0\leq{j} <2{k}}
exp(-\pi j/ {k})=0 $ and hence  these  contribution vanish when summed
over $j$. 

In section 6 only  terms of order
$q^0$ in \expanq\ will directly related to quantities in  type I$^\prime$
QM, which corresponds to taking  the limit $U_2\to \infty$.  For these
  terms the nonholomorphic
pieces in the  $(\tr(F^2))^2$ and $(\tr(R^2))^2$ due to the presence
of $\hat{E}_2$  will not supressed by inverse powers  of $U_2$ in the
$U_2\to \infty$ limit as explained in the appendix.

Using (A.5) the $I_{0,0}$ part of the integral \unfoldedint\ can be
expressed as,
\eqn\qzerores{\eqalign{I_{0,0}&= \sum_{k,p}\Big\{c^{(1)}_{0,0}{1\over
      2|p|}e^{2\pi 
    i 2kp T}+c^{(2)}_{0,0}{k+1\over (2k+1) |p|}e^{2\pi
    i (2k+1)p T}+c^{(3)}_{0,0}{1\over 2|p|}e^{2\pi
    i 2kp T}\cr &+c^{(4)}_{0,0}{k\over (2k+1) |p|}e^{2\pi
    i (2k+1)p T}\Big\} +cc.}}
In all examples considered below we find that
$c^{(2)}_{0,0}=c^{(4)}_{0,0}$, Hence  the contributions of the $A^{(2)}$ and $A^{(4)}$ sector can
    be combined, rearranging the summation gives   
\eqn\qzeroresa{I_{0,0}= \Big({ c^{(1)}_{0,0}+c^{(3)}_{0,0}\over 2}-
    c^{(2)}_{0,0}\Big)\sum_{N|n} {1\over n}e^{2\pi i2 N T}+
    c^{(2)}_{0,0} \sum_{N|n} {1\over n}e^{2\pi i N T}+cc.}
Where ${N|n}$ denotes the set of all integers $n$ which
divide $N$.
\subsec{$t_8 \tr(R^4)$ thresholds} For the $t_8 \tr(R^4)$
threshold, the operator $Q$ does not depend on the spin structures  of the
theta functions associated with the first factor $SO(16)$,
\eqn\rfourthr{Q_{R^4}= {1\over 2^7 3^2 5}{1\over
\eta^{24}(\tau)} E_4(\tau).} 
Combining \rfourthr\  with  \intlat\  $QC(A^{(i)})$ for $\tr(R^4)$ is
given by
 \eqn\chargea{\eqalign{
QC(A^{(1)})&= {1\over 2^7 3^2 5}{E_4 \over \eta^{24}}\big(\theta_2^{16}
  +\theta_3^{16}+ \theta_4^{16}\big),  \cr
QC(A^{(2)})&= {1\over 2^6 3^2 5}{E_4\over \eta^{24}}\theta_2^8 \theta_3^8
\cr QC(A^{(3)})&= {1\over 2^6 3^2 5}{E_4\over  \eta^{24}}\theta_3^8
\theta_4^8, \cr
 QC(A^{(4)})&= {1\over 2^6 3^2 5}{E_4\over\eta^{24}}\theta_2^8 \theta_4^8
.}}
Expanding the terms in \chargea, confirms that
    $c^{(1)}_{-1,0}=c^{(3)}_{-1,0}=1/2^63^2 5$ and
    $c^{(2)}_{-1,0}=c^{(4)}_{-1,0}=0$ 
    and hence  terms of order $1/q$ do vanish in the integral after
    summation over $j$. Furthermore one finds 
    $c^{(1)}_{0,0}=744/2^6 3^2 5 $,
    $c^{(2)}_{0,0}=c^{(4)}_{0,0}=256/2^6 3^2 5$ and $ 
    c^{(3)}_{0,0}=232/2^6 3^2 5$. Plugging these coefficients into
    \qzeroresa\ gives 
\eqn\trrfres{I^{\tr(R^4)}_{0,0}={1\over 2^6 3^2 5}\Big\{256\sum_{N|n}
  {1\over n}e^{2\pi i N T}+232 \sum_{N|n} {1\over n}e^{2\pi i2 N
    T}\Big\}.} 

 \subsec{$t_8\tr(F^4_1)$ thresholds}
 The operator  $ Q$ for  the threshold for $t_8\tr(F^4_1)$
 associated
to a complex fermion with spin structure $[a,b]$ was defined in \qresult.
 Using \intlat\ and \qresult\ we can
express $QC(A^{(i)})$ for the $\tr(F_1)^4$ threshold as
\eqn\charge{\eqalign{Q C(A^{(1)})&= {1\over  2^8 3}{1\over
 \eta^{24}}\big(-\theta_2^{16} 
  \theta_3^4\theta_4^4 +\theta_3^{16}
  \theta_2^4\theta_4^4- \theta_4^{16}
  \theta_2^4\theta_3^4\big)= 1,  \cr
Q  C(A^{(2)})&= {1\over  2^8 3}{1\over  \eta^{24}}\theta_2^8 \theta_3^8(
  -\theta_3^4\theta_4^4 +\theta_2^4\theta_4^4)
  = -{1\over 3}, \cr
Q  C(A^{(3)})&= {1\over  2^8 3 }{1\over \eta^{24}}\theta_3^8 \theta_4^8(
  \theta_2^4\theta_4^4 -\theta_2^4\theta_3^4)
  =- {1\over 3}, \cr
Q  C(A^{(4)})&= {1\over 2^8 3}{1\over \eta^{24}}\theta_2^8 \theta_4^8(
  -\theta_3^4\theta_4^4 -\theta_2^4\theta_3^4)
  = -{1\over 3}.}}
Where  the following  identities were used 
\eqn\identone{\theta_2^4+\theta_4^4-\theta_3^4=0,\quad
 \theta_2^4\theta_3^4\theta_4^4=16\eta^{12}, \quad 
  \theta_3^{12}-\theta_2^{12}-\theta_4^{12}=48\eta^{12}.}
Note that in \charge\ all dependence on powers of
$q^n$ with $n\neq 0$ has disappeared. With  $c^{(1)}_{0,0}=-1$ and
 $c^{(2)}_{0,0}=c^{(3)}_{0,0}=c^{(4)}_{0,0}=1/3$ the result for the
 non degenerate orbit is given by
\eqn\ndegres{I^{\tr(F^4)}_{0,0}= - {1\over 3} \sum_N\sum _{N|n} {1\over
 n} e^{-2\pi 
i N T}+{2\over 3} \sum_N\sum _{N|n}
 {1\over n} e^{-2\pi i2 N  T}+c.c.}

\subsec{$(tr(F^2)_1)^2$ thresholds}
The operator $Q$ for  the $(tr(F^2)_1)^2$ threshold
depending  on the spin structures was defined in  \trftwoq. Together
with \intlat\ $QC(A^{(i)})$ become
\eqn\ctrftwo{\eqalign{Q C(A^{(1)})&=
{1\over 2^{10}3^2}{1\over
\eta^{24}}\Big\{\theta_2^{16}\big(e_2+\hat{E}_2\big)^2
  +\theta_3^{16}\big(e_3+\hat{E}_2\big)^2
  +\theta_4^{16}\big(e_4+\hat{E}_2\big)^2\Big\}, \cr
Q  C(A^{(2)})&={1\over 2^{10}3^2} {1\over \eta^{24}}\theta_2^8
\theta_3^8\Big\{\big(e_2+\hat{E}_2\big)^2+\big(e_3
+\hat{E}_2\big)^2\Big\},
   \cr
Q  C(A^{(3)})&= {1\over 2^{10}3^2}{1\over \eta^{24}}\theta_3^8
\theta_4^8\Big\{\big(e_3+\hat{E}_2\big)^2+\big(e_4+
\hat{E}_2\big)^2\Big\},\cr 
Q  C(A^{(4)})&= {1\over 2^{10}3^2}{1\over\eta^{24}}\theta_2^8
\theta_4^8\Big\{\big(e_2+\hat{E}_2\big)^2+\big(e_4
+\hat{E}_2\big)^2\Big\}.}}
Expanding the terms in  \ctrftwo\ it is easy to confirm that
there are no  terms of order $1/q$ present.
 Furthermore we get
    $c^{(1)}_{0,0}=1/8$,$c^{(2)}_{0,0}=c^{(4)}_{0,0}=1/4$ and $
    c^{(3)}_{0,0}=1/8$ and the result for the integral is then given by 
\eqn\Inresf{I=-{1\over 8}  \sum_N \sum_{N|n} {1\over n}e^{-2\pi i
2NT}+ {1\over 4} \sum_N \sum_{N|n}{1\over n} e^{-2\pi iN T}+cc. }

\newsec{type I$^\prime$ Quantum mechanics}
In order to determine the existence of bound states of $D0$ branes an
index of the D0 brane QM has to be computed. In the case of type IIA
D0 branes in ten dimensions this was done for the case of two D0
branes in \sethi,\yi. In this calculation the index is split into a
bulk and boundary term which in turn is expressed as a zero
dimensional matrix integral.

The Hamiltonian for the $(0,8)$ quantum mechanics governing
D0-particles in type I$^\prime$  is given by (in the gauge $A_0=0$ and
following the notation of \bgs) 
\eqn\qmham{\eqalign{H&= {1\over 2}\tr \Big( \Pi_j^2-\Pi_\phi^2 +
    g^2[\Phi,X_i]^2  -{g^2\over 2}
  [X_i,X_j]^2\Big) + {ig\over
    2}\tr\Big(\lambda_{\dot{\alpha}}[\Phi,\lambda_{\dot{\alpha}}]+
  \Theta_{\alpha}[\Phi,\Theta_{\alpha}]\cr
&-2 X_i \gamma^i_{\alpha\dot \alpha} \{
\Theta_\alpha,\lambda_{\dot{\alpha}}\} \Big) -ig \Big(\chi_I^t
\Phi \chi_I+ m_{IJ}\chi^t_I\chi_J\Big). }}
 All fields but the
$\chi$ are given by an orientifold projection of the $SU(N)$
D0-particle quantum mechanics, where $X_i$ and $\Theta_\alpha$
transform as the traceless symmetric representation of $SO(N)$
which is given by the real matrices of the Lie algebra of $SU(N)$.
The spinor $\Theta_\alpha$ transforms as $8_c$ spinor of $SO(8)$
realated to  the supersymmetries of the D0 brane unbroken by the
presence of the D8 brane. In addition we have the trace part $x_i$
and $\theta_\alpha$ which are singlets under $SO(N)$ and do not
enter in the interacting Hamiltonian \qmham. $\Phi$ and
$\lambda_{\dot{\alpha}}$ transform under the adjoint
representation of $SO(N)$ which is given by the imaginary elements
of $SU(N)$ and transform as $1$ and $8_s$ of $SO(8)$ respectively.
The chiral fermions $\chi^I_i$ transform in the real $(8,2N)$ of
$SO(8)\times SO(2N)$. Giving nonzero values to the parameters
$m^{IJ}$ corresponds to moving the D8 branes away from the
orientifold planes. The index calculated below will in principle
depend on the values of the parameters $m_{IJ}$. In the following
we will mostly be interested the case of all $m^{IJ}=0$ in the
Hamiltonian. The Gauss constraint is given by 
\eqn\gaussc{G=
[\Pi_j,X_j]-[\Pi_\phi,\Phi]+i \Theta_\alpha\Theta_\alpha-i
  \lambda_{\dot{\alpha}} \lambda_{\dot{\alpha}}+i \chi_I\chi_I^T.}
The index of QM is given by 
\eqn\indexone{I_N= \lim_{\beta \to
\infty}\tr(-1)^F e^{-\beta H},} 
where the trace is taken over gauge
invariant states which satisfy $G=0$. An integration by parts
turns the index into a bulk $Z_N$ and deficit term $\delta I_N$,
where $I=Z_N+\delta I_N$ and the bulk term is given by
\eqn\indextwo{Z_N= \lim_{\beta \to 0 }\tr(-1)^F e^{-\beta H}.}

\newsec{type I$^\prime$ interpretation}
We want to use the results for the heterotic thresholds to determine
matrix integrals of D0 particles in type I$^\prime$ quantum mechanics.
For simplicity we  will consider a square torus with radii $R_1,R_2$ ,
the Kahler and complex structure moduli are then given
by
\eqn\kahlcom{T= B^{NS}_{12}+i R_1R_2,\quad U= i {R_2\over R_1}.}
Under heterotic type I duality the coupling constants, metric and AST
field  are related by
\eqn\hettypi{\lambda_{het}=1/\lambda_I,\quad  \lambda^I
  g_{\mu\nu}^{het}=g^I_{\mu\nu},\quad B_{\mu\nu}^{het}=B_{\mu\nu}^I.}
 Under a T-duality along the first circle   type $I$ gets
mapped to type I$^\prime$, where the radii are related  by
\eqn\tdual{R_1^I =1/ R^{I^\prime}, \quad R_1
  \lambda^{I^\prime}=\lambda^I,\quad B^{RR,I}_{12}=A_2^{RR,I^\prime}. }
 Hence the heterotic moduli T and U get
mapped to
\eqn\typeonep{T= A_2^{RR} +i {R_2^\prime\over \lambda^\prime},\quad
  U=i R_1^\prime 
  R_2^\prime.} 
In the type $I^\prime$ variables the heterotic modulus $T$ has the
interpretation of the action of a Euclidean D0 brane worldline on a circle
of radius $R^2$. On the other hand the modulus $U$ is independent of the
type $I^\prime$ coupling constant and can be interpreted as the action of
open string worldsheet instantons which stretch between the two
8-brane/orientifold planes. In the limit of infinite separation of the
8-brane/orientifold planes $R_1\to \infty$ all contributions of the form
$\exp(2 \pi k U)$ will therefore vanish for $k>0$.  For this limit to
be meaninful it is 
important that terms of order $1/q$ in the integral \unfoldedint\  do not contribute as mentioned in
section 4, since they will behave as $\exp(2\pi U_2)$ which diverges
as $U_2\to \infty$. 
 Note that the in the limit $U_2\to \infty$ the two O8
 planes effectively decouple. A calculation as in section 4 for a
 $t_8\tr(F_1)^2\tr(F_2)^2$ threshold, where the two traces are over
 the two different $SO(16)$, reveals that there are no terms which
 survive the $U_2\to \infty$ limit. Hence the traces involving only
 one $SO(16)$ factor should be sensitive only to the QM of D0
 particles on one O8 plane. The situation for the gravitational
threshold might be more complicated allthough the counting of
fermionic zero modes suggests that the thresholds are only related 
to D0-branes on one of the two O8 planes.
In the following we will identify
 $\exp(2\pi i N T)$ term in the threshold with the euclidean
 action for a worldline of a bound 
 state of $N$ D0 branes. The prefactor of the 
 threshold should then be related to  the bulk partition
 function $Z_N$ of the index for $N$ type $I^\prime$
 D0-particles.
This is the same idea used in \greenguta\ for the IIA D0 particle
 quantum mechanics.  For odd $N$ the $I_{0,0}$ calculated in
 \trrfres,\ndegres\ and \Inresf\ are of the same form and  it is
 tempting to conjecture that  
the value of $Z_N$ up to an $N$ independent numerical factor is given by
\eqn\zodd{Z_{N}= const \sum_{N|n}{1\over n}, \quad N \;\; odd.}
The value of $Z_N$ is determined up to an $N$ independent constant
 which  can in principle be determined by a careful analysis of the
 relative normalization of the heterotic calculation and the type
 $I^\prime$ QM. 
On the other had for even $N$, i.e. $N=2 N^\prime$, the
 results for the integrals $I_{0,0}$ have a different structure than  \zodd.
The  terms in $I_{0,0}$  \trrfres,\ndegres\ and \Inresf\ proportional
 to $\exp(2\pi i 2N^\prime T)$ are given by  
\eqn\zmurfour{Z^{tr(R^4)}_{2N^\prime}=
 {1\over 2^6 3^2 5}\Big(232\sum_{N^\prime|n}{1\over n}+256\sum_{2N^\prime|n}{1\over
 n}\Big),} 
and 
\eqn\zmulfour{Z^{tr(F^4)}_{2N^\prime}=
 {1\over 3}\Big(2\sum_{N^\prime|n}{1\over n}-\sum_{2N^\prime|n}{1\over
 n}\Big),} 
and 
\eqn\zmulfourb{Z^{(tr(F^2))^2}_{2N^\prime}=
 {1\over 8}\Big(-\sum_{N^\prime|n}{1\over n}+2\sum_{2N^\prime|n}{1\over
 n}\Big).} 
All these expression are of the form
\eqn\zmulfoura{Z_{2N^\prime}=
 c_1\sum_{N^\prime|n}{1\over n}+c_2\sum_{2N^\prime|n}{1\over
 n}.} 
with some constants $c_1$ and $c_2$ and it is natural to assume that the bulk
 part of the index has the same structure, allthough it is at present not 
clear whether one can read off the value of the
 constants  from \zmurfour\ directly.

A possible explanation for this behavior of the thresholds for even
 $N$  could be  that the 
 heterotic threshold corrections  are not
 related directly to  the bulk part of the index but to 
 some correlation function for the QM, which
 differs  from    the bulk index for $2N$ D0 particles
 but is proportional to  it for odd
 number of $D0$ particles.

The fields entering the quantum mechanics also contain eight
singlet bosons $x_i.i=1,\cdots,8$ and eight fermions
$\theta_a,a=1,\cdots 8$. The $R^4$ threshold  then corresponds
to the loop amplitude of a D0 brane coupling to four gravitons.
The vertex operator for a graviton with polarization ternsor $h_{ij}$ 
coupling to the D0 branes is
given by \eqn\gravvertc{V(h)= h_{ij}k_k
\theta^\alpha\gamma^{ik}_{\alpha\beta}\theta^\beta \dot{X}^j
e^{ikX}. }
 The insertions of the four graviton vertices soaks up the eight
fermionic zero modes $\theta_a$. The $SO(N)$ part of the QM does not
couple to these `center of 
mass' coordinates and hence the $R^4$ threshold should then  be multiplied by a
partition function of the $SO(N)$ degrees of freedom which we
interpret as the bulk term for the QM. 

In the case of $\tr(F^4)$ and $(\tr(F^2))^2$, the situation is more
complicated since the gauge fields  live on the D8 
brane and there is a coupling between these and the $SO(N)$ QM via
$\chi_I^a$. A vertex for a gauge field with field strength
$F_{ij}^{IJ}$ for a D0 brane is of the 
form \eqn\gaugcoup{V(F)= F_{ij}^{IJ}\theta^\alpha
\gamma^{ij}_{\alpha\beta}\theta^\beta \chi^I_a\chi^J_a e^{ikX}.}
Hence a D0-brane loop coupling to four gauge fields will
correspond to the insertion of four such vertex operators, which
again soak up the eight $\theta$ zero modes. The insertion of
$\chi^I_a\chi^J_a$  in the path integral  of the $SO(N)$ is
equivalent to taking derivatives
 $\partial/\partial m_{IJ}$ of $Z_N$   in \indextwo.
The results of the heterotic threshold corrections predict these
correlation functions, if this interpretation is correct.
\eqn\qmcorf{\eqalign{(\tr(F^2))^2&:\quad{\partial\over
\partial{m_{IJ}}}{\partial\over
\partial{m_{IJ}}}{\partial\over \partial{m_{KL}}}{\partial\over
\partial{m_{KL}}}Z_N\Big|_{m_{AB}=0},\cr
 \tr(F^4)&:\quad{\partial\over
\partial{m_{IJ}}}{\partial\over 
\partial{m_{JK}}}{\partial\over \partial{m_{KL}}}{\partial\over
\partial{m_{LI}}}Z_N\Big|_{m_{AB}=0}.}}
In particular the heterotic threshold calculation implies that
there  is a difference between the case $N$ even and $N$ odd for
these correlation functions. It would be interesting to check this
conjectured result explicitly.

\newsec{D0-particle loop}
There is a simple picture   of the result for the bulk part
of the index $Z_N$ in \zodd. $N$ D0 particles will form bound
states which transform (according to heterotic type I duality) as
the {\bf 128} of $SO(16)$ for odd $N$ and as the {\bf 120} of
$SO(16)$ for even N. One can imagine that the D0 particles are
stuck on the D8-O8 branes. The $\tr(F_1)^4$ can then be
interpreted as coming from a loop of D0 particles with four
graviton vertex operators \gravvertc\ inserted.

\eqn\looponea{I= \sum_n{1\over \pi^{D/2}}\int d^Dp \int {dt \over t}
  t^k \exp\big( -t(p^2+ 
  \mu^2 -{(n-A)^2\over R^2}
\big),  } 
where $\mu$ is the mass of a D-particle. If the D-particle is stuck on
  the D8-brane the momentum 
  integral is nine dimensional, i.e. $D=9$. Furthermore inserting
  four vertices to soak up fermionic zero modes introduces a factor of
  $t^4$, i.e. $k=4$ in \looponea. After integrating out the loop
  momentum and performing a Poisson resummation over $n$,  we get
\eqn\looptwo{\eqalign{I&= R\sqrt{\pi}\sum_m\int {dt \over
  t^{3\over 2}}\exp\big( -{\pi^2 R^2 m^2\over 
  t}- \mu^2  t+2\pi i m A \big)\cr
&=  \sum_m{1\over\  m }\exp\big( - 2\pi R m \mu+2\pi i m A \big).
}} 
In the second line formula (A.3) from the appendix has been used.
The mass $\mu$ of a D0 particle of  charge $n$  given by $\mu =
  n/\lambda$  and  $A=n A^{RR}$. Hence summming  over  the
  contribution of a charge $n$ D0 particle 
  winding $m$ times givs
\eqn\zchwin{Z=\sum_{m,n} {1\over m}e^{-2\pi mn R/\lambda+2\pi i mn A
  }= \sum_N\sum_{N|n} {1\over n} e^{-2\pi N R/\lambda+2\pi i N A}.}
Note that there is a differences to the case of the IIB D0
  particle analysis given in \greenguta. Due to the fact that the
  momentum integral is only nine dimensional (since we assumed that
  the D0 brane is stuck on the orientifold plane) the integral reduces to
  a Bessel function $K_{1/2}$ instead of $K_1$. Since the series expansion for
  $K_{1/2}$ terminates after one term  this implies  that there
  is no infinite asymptotic series of corrections. This behavior should reflect the exact cancellation of
  bosonic and fermionic fluctuations for the type $I^\prime$
  quantum mechanics.  Note also that the extra contribution for the
  $\tr(F)^4$ and $(\tr(F^2))^2$ thresholds for $N=2N^\prime$ could  be
  interpreted as 
  coming from $N^\prime$ D0 particles with charge two. This might come
  from D0 particle pairs which move pairwise off the D8 plane and form
  a bound state with twice the charge.

\newsec{Conclusions}
In this note a heterotic one loop calculation of threshold corrections
in the presence of Wilson lines was performed.   Using the heterotic
type I duality and the T-duality relating type I and type I$^\prime$
it was argued that from  these thresholds information about certain
quantities calculated in type I$^\prime$ quantum mechanics can be
extracted. At the
moment the status of this claim is not certain. A puzzling feature is
the difference in the structure of $Z_N$ calculated in section 4  for
even and odd number of D0 particles.  In particular for the
gravitational threshold $I^{\tr(R^4)}_{0,0}$ seems unlikely to be
directly related to the bulk term for even $N$. One (disappointing)
possibility is that the quantities calculated in this note are not
directly related to the bulk terms of the index (or correlation functions) for  type I$^\prime$
QM. On the other hand it would 
be very interesting to adress this question by a direct calculation of
in  type I$^\prime$ QM for arbitary $N$. This seems to be a very
difficult task. Another interesting question would be to consider more
general Wilson lines  than \wilsonl\ corresponding to moving the
D8 branes off the orientifold planes \bgs\berg. In principle the 
calculation in this note  can easily be generalized to the more general case.
 In addition it is not clear wether the part of the threshold which
depend on $\exp(2\pi i U)$  have an interpretation
in the D0-brane quantum mechanics. 
\medskip
\noindent{\bf Acknowledgments}
\medskip
I wish to thank   Piljin Yi for useful correspondences and
comments.  I am  grateful to Stephan Stieberger for pointing out an
error in a previous version of the paper. 
This work was supported in part by NSF grant PHY-9802484. 
\medskip

\appendix{A}{Evaluation of the integrals}
In this appendix we review the evaluation of integrals appearing
in the heterotic threshold calculation. The basic technique was
developed in \dixon\ for more details in this context see
\costas\kiritsis.

\eqn\intega{I_{n,r}= \sum_{k>0, 0\leq j<k,  p\neq 0}e^{2\pi i kp T}\int {d^2\tau\over \tau_2^2}
\exp\Big( {\pi T_2\over \tau_2 U_2}|k\tau+j+pU|^2\Big) {1\over
\tau_2^r} \exp (2\pi i \tau n).} 
Integrating over $\tau_1$ gives
\eqn\inetgb{\eqalign{I_{n,r}&=\sum_{k>0, 0\leq j<k,  p\neq 0}{\sqrt{U_2}\over k\sqrt{T_2}} e^{2\pi i kp
T}e^{2\pi i n (j+pU_1)/k+2\pi kp T_2}\cr
&\times\int {d\tau_2\over
\tau_2^{3/2+r}} e^{-{\pi T_2\over U_2}(k+{nU_2\over kT_2})^2\tau_2
}e^{-\pi p^2 T_2U_2/\tau_2}.}}
 The integral over $\tau_2$ can be
done using the formula \eqn\itttwo{\int_0^\infty {dx\over
x^{3/2+r}} e^{-ax-b/x}=\left(-{d\over db}\right)^r
\sqrt{{\pi\over b}}e^{-2\sqrt{ab}},}
 where 
\eqn\abdef{a= {\pi T_2\over
U_2}(k+{nU_2\over kT_2})^2,\quad b= \pi p^2 T_2U_2.}
 We are primarily interested in the evaluation of the integrals in the
large $U_2$ limit. it is easy to see that the leading contribution
in \itttwo\ is obtained when all the derivatives act on the
exponential in \itttwo. Since $a= \pi n^2U_2/(T_2k^2)+o(1)$ all
the integrals $I_{0,r}$ will be suppressed by factors of $1\over
U^r_2$ for $r>0$, because the leading term is proportional to $n$,
which vanishes for $n=0$.
The final result for for the leading $U_2$
independent term in $I_{0,0}$  is the given by
\eqn\finintn{I_{0,0}= \sum_j\sum_{k>0,p>0}{1\over k|p|} e^{2\pi i kp T}+cc. }
Where in applications of this formula in  section 4 the sumation range
of $j$ depends on the sector $A^{(i)}$ which is considered.

 \listrefs

\end